\documentclass[11pt,twoside]{article}
\usepackage{hepro5}
\usepackage{graphicx}

\usepackage[T1]{fontenc} 

\usepackage{latexsym}
\usepackage{verbatim}
\usepackage{natbib}

\newcommand{\jd}[1]{\,{\rm #1}}

\begin{document}

\vskip 1.0cm
\markboth{Sukov\'a et al.}{Chaos in accreting black holes}
\pagestyle{myheadings}

\vspace*{0.5cm}
\title{Numerical test of the method for revealing traces of deterministic chaos in the accreting black holes}

\author{P.~Sukov\'a$^1$ and A.~Janiuk$^1$}
\affil{$^1$Center for Theoretical Physics, Polish Academy of Sciences, Al. Lotnik\'ow 32/46, 02-668 Warszawa, Poland}

\begin{abstract}
The high energy radiation emitted by black hole X-ray binaries originates in an accretion disk, hence the variability of the lightcurves mirrors the dynamics of the disc. 
We study the time evolution of the emitted flux in order to find evidences, that low dimensional non-linear equations govern the accretion flow.
Here we test the capabilities of our novel method to find chaotic behaviour on the two numerical time series describing the motion of a test particle around a black hole surrounded by a thin massive disc, one being regular and the other one chaotic.

\end{abstract}

\section{Introduction}
In this paper we test our method for revealing the traces of non-linear dynamics in the observed X-ray lightcurves on two numerical trajectories, from which one is regular and the other one is chaotic. We developed and described in details the method in the paper \cite{velkyChaos}. 
We also refer the reader to the article by A. Janiuk et al in these proceedings to learn more about astrophysical applications of this method. Here we only briefly summarize its key features.

We compute the estimate of R\'enyi's entropy $K_2$ \citep{Grassberger1983227} using the recurrence analysis\footnote{Using software package provided at \texttt{http://tocsy.pik-potsdam.de/commandline-rp.php}.} for the time series. This value is compared with the values obtained for $N^{\rm surr} = 100$ surrogates made in such a way, that they share the value distribution and power spectra with the original series (IAAFT surrogates). We use the software package TISEAN \citep{1999chao.dyn.10005H,Schreiber2000346}. The significance of the non-linearity  is given by 
\begin{equation}
\mathcal{S}(\epsilon) = \frac{N_{\rm sl}(\epsilon)}{N^{\rm surr}} \mathcal{S}_{\rm sl} - {\rm sign}( Q^{\rm obs} (\epsilon) - \bar{Q}^{\rm surr}(\epsilon) ) \frac{N^{\rm surr} - N_{\mathcal{S}_{\rm sl}} (\epsilon)}{N^{\rm surr}}  \mathcal{S}_{K_2}(\epsilon) , \label{significance}
\end{equation}
for chosen recurrence threshold $\epsilon$, where $N_{\rm sl}$ is the number of surrogates, which have only short diagonal lines in their recurrence matrix, $Q^{\rm obs}$ and ${Q}^{\rm surr}_i$ are the natural logarithms of $K_2$ for the observed and surrogate data, respectively, $\bar{Q}^{\rm surr}$ is the averaged value of the set ${Q}^{\rm surr}_i$, $\mathcal{S}_{\rm sl} =3$ and  $\mathcal{S}_{K_2}$ is the significance computed from surrogates with enough long lines according to the relation
\begin{equation}
\mathcal{S}_{K_2} (\epsilon) = \frac{| Q^{\rm obs} (\epsilon) - \bar{Q}^{\rm surr}(\epsilon) |}{\sigma_{Q^{\rm surr}(\epsilon)}}.
\end{equation}

We quantize the results with $\bar{\mathcal{S}}_{K_2}$ -- the average of ${\mathcal{S}}_{K_2}$ over a range of $\epsilon$.

\section{Testing the method with simulated time series}
\label{sect:poincare}

\begin{figure*}
\includegraphics[width=\textwidth]{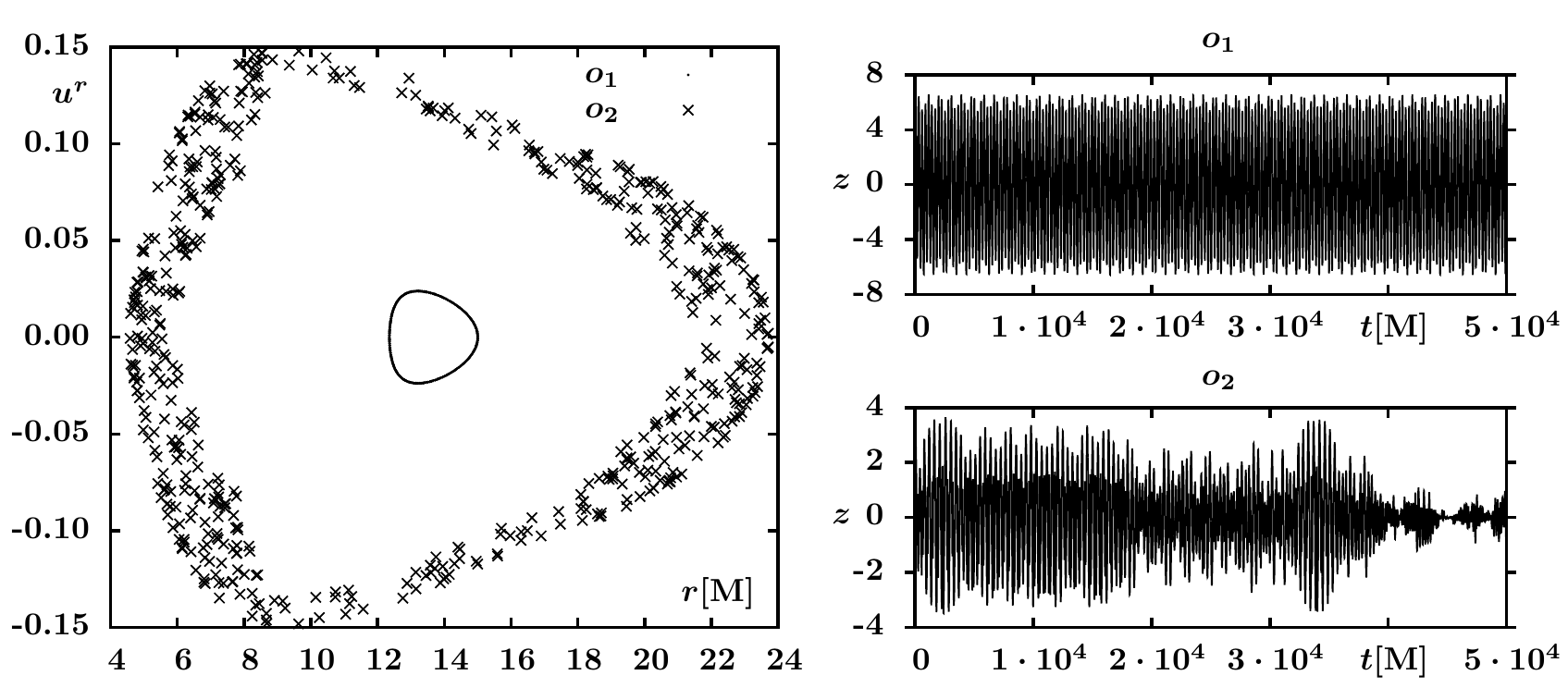}
\caption[C.1]{ Poincar\' e surface of section of regular orbit $o_1$ and chaotic orbit $o_2$ (left) and the corresponding time dependence of $z$ coordinate of these orbits (right).
}
\label{fig:Poincare}
\end{figure*}

In general our method can be applied to different kinds of time series, which are produced by some dynamical system. 
Here we test the method applying it on time series, whose nature is known. 
We choose the numerical time series, which describe the motion of geodesic test particle in the field of a static black hole surrounded by a massive thin disc. 
The background metric is given by an exact solution of Einstein equations and is described in details in \cite{semerak2010free}. The time series are obtained as the numerical solution to the geodesic equation with this metric using the 6th order Runge-Kutta method 
As the input data we use the time dependence of the particle's  $z$-coordinate.

We study two numerical trajectories, orbit $o_1$ being regular and orbit $o_2$ chaotic,
whose Poincar\'e surface of section\footnote{Poincar\'e surface of section is a method for visualization of the features in the phase space of dynamical system, which is very useful for low-dimensional systems. 
The surface of section is a chosen surface in the phase space, on which the intersections of the phase trajectory are plotted. The regular orbits draw a closed curve whereas chaotic orbits fill some non-zero area. More details can be found e.g. in \cite{semerak2010free}.} and time dependence of $z$ coordinate is depicted in Fig.~\ref{fig:Poincare}.
The two selected trajectories belong to the regular island ($o_1$) and chaotic sea ($o_2$) depicted in Fig. 19 of \cite{witzany2015free}. 

We sample the trajectory with ${\rm d}\tau=10\jd{M}$ for $\tau_{\rm max}=50\,000\jd{M}$ yielding the data set of $N=5000$ points. 
The first minimum of mutual information is at $\Delta t =90\jd{M}=k \Delta \tau, k=9$ for $o_1$ and $\Delta t =110\jd{M}, k=11$ for $o_2$, hence we adopt $\Delta t = 100\jd{M}, k=10$ for both orbits.
We generate the set of surrogates and perform the analysis in the same way as for the observed X-ray lightcurves in \cite{velkyChaos}.

At first we investigate the dependence of the length of the longest diagonal line present in the recurrence matrix $L_{\rm max}$  on $\epsilon$.
 As expected, the regular trajectory yields very long diagonal lines for small thresholds and $L_{\rm max}$ goes up almost to the maximal value $N$. 
The surrogates behave in a similar way for a little bit higher threshold. 
This is due to the way how the surrogate data are constructed, as they have exactly the same value distribution but they reproduce the spectrum only approximately depending also on the available length of the data set. In case of regular motion, very narrow peaks are in the spectrum and the error in reproducing such spectrum causes the very long diagonal lines to be broken. 
This higher value of $\epsilon$ for surrogates corresponds to the size of the neighbourhood needed for covering the small discrepancies of the surrogates. 
The chaotic orbit $o_2$ provides shorter lines, so that $L_{\rm max}< 2000$  for the range of thresholds we used.  
 Yet it is significantly larger than the corresponding values for surrogates. Only for very high thresholds, the difference decreases. 

\begin{figure*}

\includegraphics[width=0.495\columnwidth]{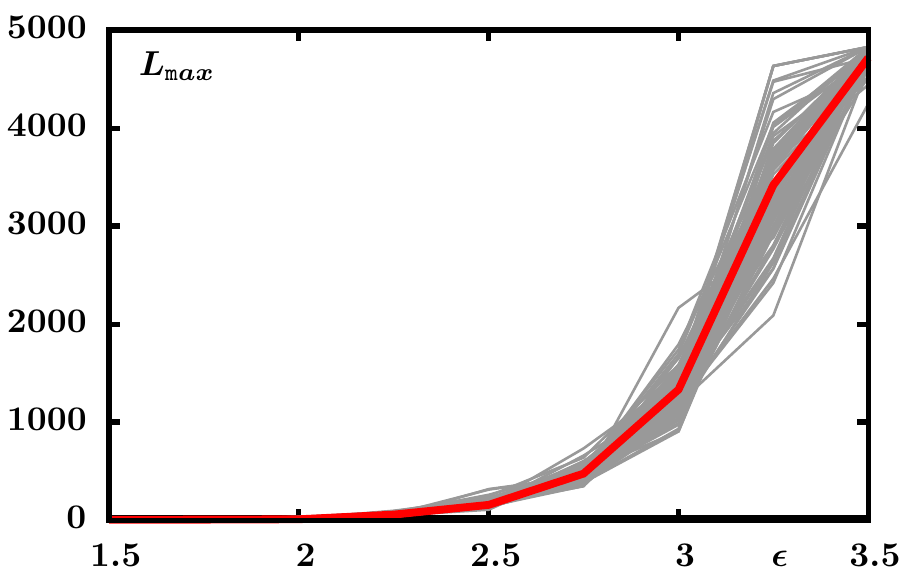}
\includegraphics[width=0.495\columnwidth]{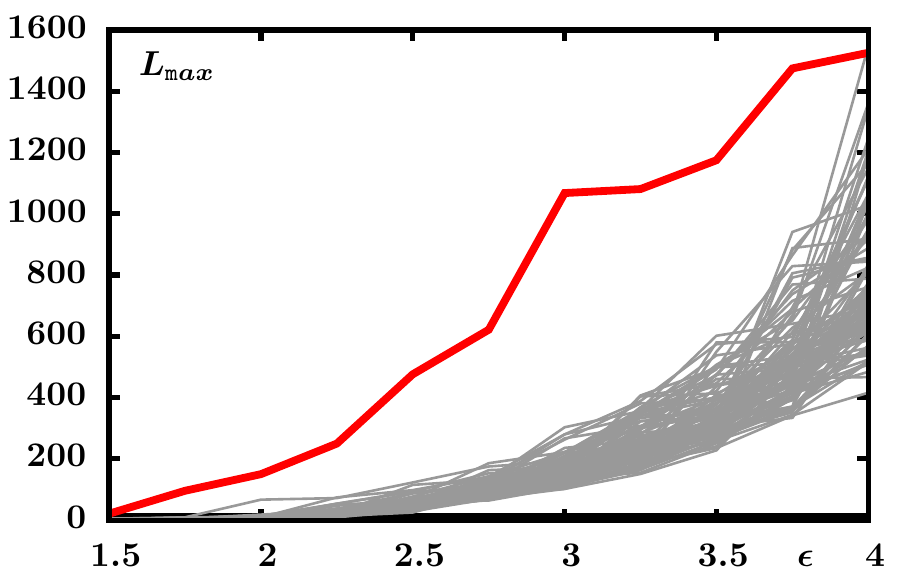}

\caption{$L_{\rm max}$ for regular orbit $o_1$ (left) and chaotic orbit $o_2$ (right) with added noise with $\sigma_n=0.4$ plotted by thick red lines and the same for the ensemble of surrogates.
}
\label{fig:Preg_Lmax}
\end{figure*}

Because in reality the data always contain some level of noise, we take the normalized times series for $o_1$ and $o_2$ and we add a white noise with zero mean and increasing variance $\sigma_n$ and rescale the resulting data back to zero mean and unit variance. 
The surrogates created from the regular orbit with added noise reproduce the spectrum better than for the regular orbit alone (normalised rms discrepancy between the exact spectrum and the exact amplitude stage reported by the {\tt surrogates} procedure decreases from $\sigma_n=0$ to $\sigma_n=0.25$).

In Fig.~\ref{fig:Preg_Lmax} the plots of $L_{\rm max}$ versus $\epsilon$ for the added white noise with $\sigma_n = 0.4$ are shown. 
The presence of noise shifts up the needed threshold for some lines to occur in RP. 
For the regular orbit $o_1$ there is no significant difference from the surrogates.
For chaotic orbit $o_2$ the threshold is also shifted to higher values, but the difference from the surrogates remains.

Our posed null hypothesis is that the data are the product of linearly autocorrelated process and because the regular trajectory can be treated as such (e.g. in the case of a periodic orbit, the points separated by the period $T$ are the same), the significance is small. On the other hand, the chaotic trajectory cannot be treated as linear dynamics and yields high significance.

\begin{figure*}
\includegraphics[width=0.495\columnwidth]{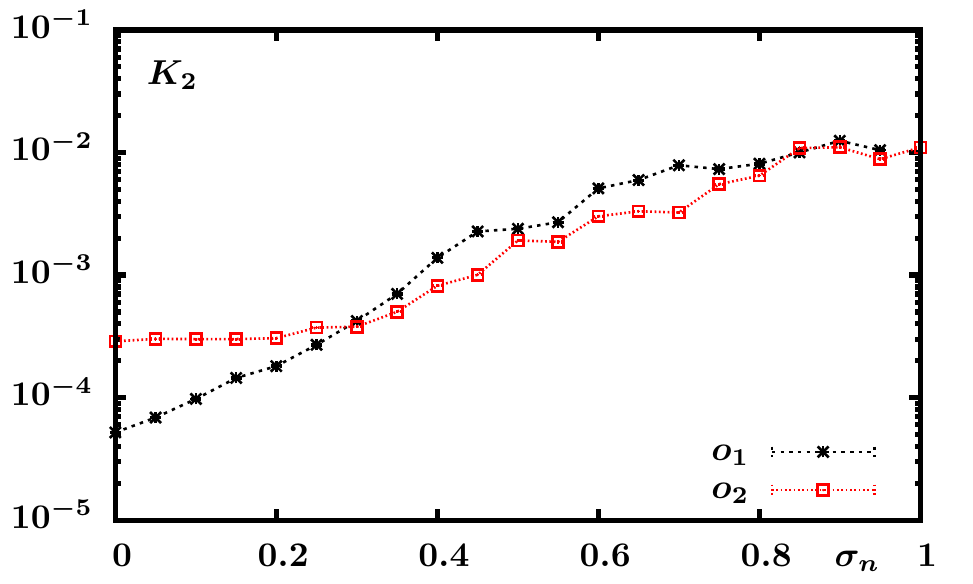}
\includegraphics[width=0.495\columnwidth]{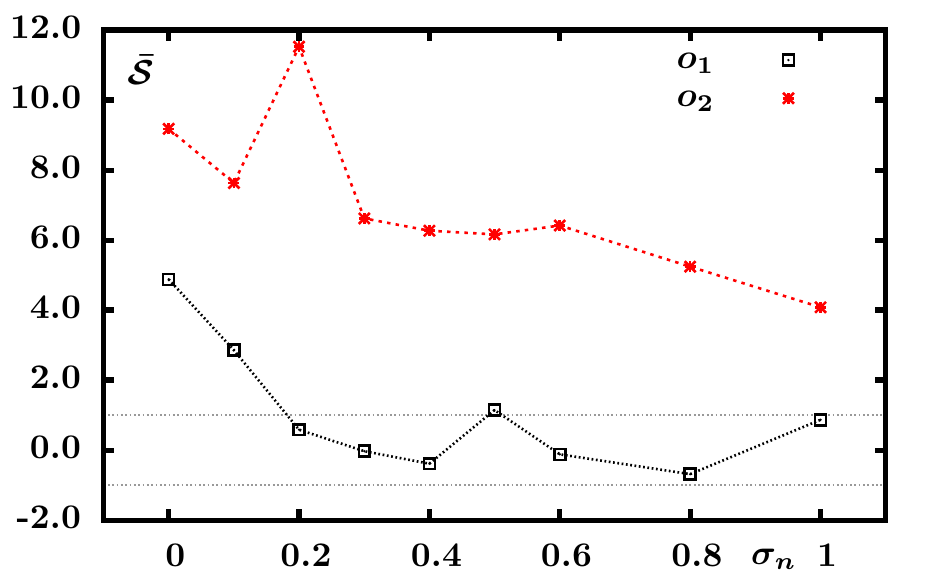}

\caption{The estimates of the R\'enyi's entropy $K_2$ computed for RR$\sim$15\% and the significance of its comparison with surrogates with increasing strength of the noise.
}
\label{fig:Preg_Lmax2}
\end{figure*}

In Fig.~\ref{fig:Preg_Lmax2} the estimate of $K_2$ and the significance of its comparison with the surrogates is given for the increasing level of noise ($\sigma_n =0, 0.05,  \dots, 1.00$). 
We note, that for low levels of noise the regular orbit yields much lower value of $K_2$, which can serve as the differentiation between chaotic and regular motion (see \citep{semerak2012free}). 
However for increasing strength of the noise, the regular orbit $o_1$ seems to be more affected than the chaotic one, providing higher $K_2$ for the noise levels $\sigma_n>0.25$. 
Therefore, the significance for the regular orbit drops down bellow one quickly, while the significance for chaotic orbit reaches values around 10 for low noise levels, near 6 for intermediate noise levels and stays around 4 for high noise levels, even up to the case, when the variance of the noise is the same as the variance of the data.

\section{Conclusions}
\label{discussion}

The test of the method for finding non-linear dynamics in dynamical systems based on the observed time series shows that for a given length of observational data set (5000 points) we can expect that chaotic dynamics would yield values of significance between 2-10 depending on the strength of the noise. Regular motion would not provide significant result, because even low level of noise present in the measured data destroys the differences with respect to the surrogates.



\acknowledgments This work was supported in part by the grant DEC-2012/05/E/ST9/03914 from the
Polish National Science Center.




\end{document}